\documentclass[letterpaper, 10 pt, conference]{ieeeconf}  

\usepackage{amsmath, amssymb}
\usepackage{amsmath}
\usepackage{algorithm}
\usepackage{graphicx}
\usepackage{multirow}
\usepackage{subfig}
\usepackage[noend]{algpseudocode}
\usepackage{etoolbox}

\IEEEoverridecommandlockouts                              

\makeatletter
\patchcmd{\fs@ruled}
  {\def\@fs@pre{\hrule height.8pt depth0pt \kern2pt}}
  {\def\@fs@pre{\vspace*{5pt}\hrule height.8pt depth0pt \kern2pt}}
{}{}
\floatstyle{ruled}
\restylefloat{algorithm}
\makeatother

\title{\LARGE \bf
Security and Performance Considerations in ROS 2: A Balancing Act
}

\author{Jongkil Kim$^{1}$, Jonathon M. Smereka$^{2}$, Calvin Cheung$^{2}$, Surya Nepal$^{1}$ and Marthie Grobler$^{1}$ %
\thanks{*This work was sponsored by the U.S. Army Tank Automotive Research, Development, and Engineering Center (TARDEC)}%
\thanks{$^{1}$Jongkil Kim, Surya Nepal and Marthie Grobler are with the Commonwealth Scientific and Industrial Research Organization (CSIRO) Data61,
        Marsfield NSW 2122, Docklands VIC 3008, Australia
        {\tt\small jongkil.kim@data61.csiro.au, surya.nepal@data61.csiro.au, marthie.grobler@data61.csiro.au} (Jongkil Kim is also affiliated with University of Wollongong.)} %
\thanks{$^{2}$Jonathon M. Smereka and Calvin Cheung are with U.S. Army TARDEC Ground Vehicle Robotics, Warren, MI 48397 USA
        {\tt\small jonathon.m.smereka.civ@mail.mil, calvin.m.cheung.civ@mail.mil}}%
}

\begin{document}

\maketitle
\thispagestyle{empty}
\pagestyle{empty}

\begin{abstract}

Robot Operating System (ROS) 2 is a ground-up re-design of ROS 1 to support performance critical cyber-physical systems (CPSs) using the Data Distribution Service (DDS) middleware. Accordingly, the security of ROS 2 is highly reliant on the security of its DDS communication protocol.  However, finding a balance between the performance and security is non-trivial task. Inappropriate security implementations may cause not only significant loss on performance of the system, but also security failures in the system. In this paper, we provide an analysis of the DDS security protocol as well as an overview on how to find the balance between performance and security. To accomplish this, we evaluate the latency and throughput of the communication protocols of ROS 2 in both wired and wireless networks, and measure the efficiency loss caused by the enabling of security protocols such as Virtual Private Network (VPN) and DDS security protocol in ROS 2 in both network setups. The result can be directly used by robotics developers to find the optimal and balanced settings of ROS  2 applications. Additionally, we analyzed the security specification of DDS using existing security standards and tested the implementation of the DDS protocol by performing static analysis. The results of this work can be used to enhance the security of ROS 2.

\end{abstract}

\section{Introduction}



The Robot Operating System (ROS) \cite{ROS2009} is one of most popular frameworks to develop robotics and drones, which are types of Cyber Physical Systems (CPSs) \cite{AnalyzeCPS}.  ROS 2 is a ground-up re-design of ROS 1 to support the expanded scope of robotics that utilize ROS. It moves away from a design for ideal conditions and move towards middleware that can support production environments (real-time systems, non-ideal networks, small embedded platforms, etc.). The most significant change in design is the abstraction of the communications middleware implementation. That is, rather than implementing communication through custom protocols (e.g., ROSTCP and ROSUDP communicating via XML-RPC), ROS 2 is built on top of existing middlewares, namely the Data Distribution Service (DDS) \cite{DDS1_1}.

The DDS protocol is developed for fast communication, and therefore latency and throughput are critical performance parameters within the service. ROS 2 performance without security (i.e., \emph{clear text} communication) is already well observed and documented by eProsima \cite{eProsima}, whilst the performance with the enabled security functions have not been thoroughly assessed. As ROS 2 matures into a product for real world applications, the CPSs that utilize ROS 2 have significant real world concerns regarding cybersecurity \cite{SPIE_ROS_Security, 2018arXiv180803322D}. Hence, it is important to understand the trade-offs and risks involved with sacrificing security for performance \cite{RobotSecuritySafety}. The Object Management Group (OMG), who maintains the standard for DDS, already considers the security of DDS by providing in-depth security specifications \cite{DDS1_1}. However, as we show in this work, there are still inconsistencies between the specification and resulting implementation in ROS 2.

This research focuses on investigating and characterizing the performance of ROS 2 with the recently implemented security mechanisms. We assess the cryptographic algorithms and their implementations using the standards derived from NIST SP 800-53 \cite{NIST800_53}, and Federal Information Processing Standards (FIPS) 140-2 \cite{FIPS140_2} and FIPS 140-2 DTR \cite{FIPS140_2DTR}. Our tests and analysis mainly focus on the correctness of the implementation and the key management aspect of the DDS protocol. In addition, we provide recommendations from our assessment such as forward secrecy. Finally, we detail aspects that can potentially make the system vulnerable, e.g., inconsistencies between the specification and implementation.

In this work we:

\begin{itemize}
\item Perform physical experiments by implementing ROS 2 to measure the latency and throughput performance of the DDS security functions.
\item Build a secure network to compare the performance effects with the use of a Virtual Private Network (VPN) versus with the performance effects of the DDS security protocol, testing in wired and wireless settings. 
\item Perform static analysis of the eProsima Real Time Publish Subscribe (RTPS) communication standard, the default middleware in ROS 2. 
\item Provide recommendations in terms of specific actions in the DDS implementation that will resolve the presented security assessment failures. 
\end{itemize}


An overview of the DDS protocol security specification is presented in Section \ref{ddsprot}. The approach used for physical analysis and implementation over both wired and wireless networks is discussed in Section \ref{approach}, while experiments and results are detailed in Section \ref{physical_analysis}. Finally, static analysis of the eProsima RTPS middleware for ROS 2 and DDS implementation recommendations are provided in Section \ref{ddsrec}. 


\section{Background}

ROS was initially designed for academic and hobbyist communities with the intent to create software tools to enable robotic development that could be easily shared and reused \cite{ROS2Design}. Accordingly, cybersecurity protections were not a focus (i.e., flawed security is worse than no security) so previous efforts sought to examine how to include security protections and their impact \cite{SPIE_ROS_Security, 2018arXiv180803322D, ROS_AppSecurity, ROS_Security_book, ROS1_Security, ROSCaseStudy, ROSForensic, SecureComsROS}.
\subsection{ROS and its Security Posture}
 
Despite the lack of built-in cybersecurity protections into ROS 1, many parties have been keenly intersted in the adapation of ROS, namely the U.S. Army, with the goal of improving the cybersecurity posture of the entire framework and to ensure it is viable for military use \cite{SignalMagazine}. The primary driver for this adaptation is Department of Defense (DoD) level guidance pushing for the use of a modular and open system approach to development and acquisition across all programs \cite{DODOSA}. Given the need for modularity, reuse, and open architectures \cite{ARMY_RAS}, ROS meets many of the needs for Army robotics development, with the important exception of cybersecurity. Due to the highly sensitive nature of information transferred and stored on DoD information systems, all CPSs that connect to the Global Information Grid must be extensively reviewed for cybersecurity risks \cite{DoD8510}. In order to leverage the modularity benefits of ROS while maintaining an acceptable level of cybersecurity risk, ROS-Military (ROS-M) is being established to have a military ecosystem for ROS 1 and 2 applications that takes advantage of current cybersecurity improvement efforts and funding further development in that domain.  

The new version of ROS aims to address the issue of cybersecurity through two core efforts; connection-oriented security and application-oriented security (i.e., access control and auditing). The focus of this work is on the implementation and performance of ROS 2 based on the developed connection-oriented security measures. Although, unlike previous work \cite{SecureComsROS, ROS1_Security, ROS_AppSecurity, SROS2016} which aims to analyze or develop methods of securing communication between ROS 1 nodes, this effort focuses on investigating the security measures built into ROS 2, which relies on third party communication middleware using DDS.

\subsection{DDS and its Security}\label{ddsprot}

The security of ROS 2 is highly reliant on the security of its DSS communication protocol; hence, our analysis focuses heavily on evaluating the implementation and efficiency of DDS in ROS 2. The DDS protocol aims to realize a fast publish-subscribe system between multiple ROS 2 nodes. Historically, the DDS specification relied on secure networks, i.e., there was no authentication, encryption, or access control measures in place. OMG has recently produced a security standard \cite{DDS1_1} to add new security compliance points to the DDS specification, however, as we show, there are still some inconsistencies between that specification and resulting applications for at least one implementation. 

The security standard for DDS implements a three-way handshake consisting of \textbf{HandshakeRequest}, \textbf{HandshakeReply} and \textbf{HandshakeFinal} messages. It identifies participants based on certificates using Public Key Infrastructure (PKI) and uses the Diffie-Hellman (DH) key exchange protocol. As an example, consider two participants where a handshake is made in the following way:
\begin{enumerate}
\item \textbf{HandshakeRequest} message is sent by Participant 1 to initiate the key agreement protocol. It contains Participant 1's certificate information, a DH public key, and a random nonce. 
\item \textbf{HandshakeReply} message is sent by Participant 2 in response to the HandshakeRequest message. It contains Participant 2's certificate information, a DH public key, Participant 1's random nonce, as well as another random nonce. The message is used to verify Participant 2's receipt since the whole message is signed by Participant 2. 
\item \textbf{HandshakeFinal} message is sent by Participant 1 to confirm the receipt of HandshakeReply. It contains both nonces and is signed by Participant 1.  
\end{enumerate}

After the handshake, the DDS protocol uses AES-GCM to encrypt and decrypt the data on a secure channel. It may also provide additional reader-specific message authentication codes using Galois MAC (AES-GMAC) \cite{NIST800_38}. The AES-GCM transformation produces both the ciphertext and a Message Authentication Code (MAC) using the same secret key. After a shared secret is established and both nonces were exchanged between two participants using the handshake protocol, participants can build master secrets using the HMAC-Based Key Derivation Function (HKDF) recommended by IETF RFC 5869 \cite{IETF_RFC_5869}. When implemented to the specification, this is sufficient to protect clear text and ensure data integrity. However, when not implemented to the specification, there is a looming question of data integrity. In this work, we show that both ROS 2 and the default DDS middleware could benefit from additional security considerations to better protect data integrity.


\section{Evaluation Framework}\label{approach}


The DDS protocol offers a local publish/subscribe system into ROS 2 applications for low latency communication. However, the performance of the DDS protocol may vary depending on its implementation. According to eProsima \cite{eProsima}, one can achieve a high throughput and a low latency, both desirable properties for the real-time operations. In this section, we provide a framework for performing the benchmark evaluation in various combinations of network and security settings to reflect the real word applications. We used two machines, called \textbf{MPub} and \textbf{MSub}. Their specifications are summarized in Table \ref{Tab1}, in the experiments in Section \ref{physical_analysis} to test throughput and latency performance. 

\begin{table}
	\centering
	\vspace{2mm}
    \caption{Specification of test machines} \label{Tab1}
	\begin{tabular}{c||c|c}
	Machine names	& \textbf{MPub} & \textbf{MSub} \\
		\hline\hline
	Host OS & Windows 10 & Windows 10 \\
	Guest OS & Ubuntu 16.04 LTS & Ubuntu 16.04 LTS\\
	Memory (Total(Guest)) & 16 GB (4GB) & 4 GB (2GB)\\
	\multirow{2}{*}{Total CPUs} & Intel i5-7740HQ & Intel N4200 \\
    &  2.80 GHz (8 Cores) &  1.10 GHz (8 Cores)\\
	Guest CPUs &  2.80 GHz (4 Cores)& 1.10 GHz (4 Cores) \\
	Installed VPN module & Client & Server
    \end{tabular}
\end{table}

\subsection{Performance Metrics}\label{perfmetrics}
To measure the communication efficiency, we define two performance metrics:

\begin{enumerate}
\item \underline{Estimated Latency} is a delay based on a round-trip time of a packet. Algorithms \ref{LPub} and \ref{LSub} show pseudocode to measure the Estimated Latency. A publisher measures the time taken from publishing the message ($T_1$) to subscribing the same message ($T_2$), and computes the latency by calculating $(T_2- T_1)/2$. The subscriber is a simple node that publishes the subscribed message to the other channel as soon as it is received. We use a round trip latency since measuring one-way latency requires two machines with highly synchronized clocks, which is quite difficult to build and verify when there is a very small latency ($<$ 1 ms).

\item \underline{Estimated Throughput} is the number of bits per second measured by a publisher. A publisher node sends a certain number of packets to a subscriber node without receiving any reply and measures the time it takes to send all those packets. After sending the packets, the publisher receives the number of successfully received packets from the sender. The throughput is then calculated using the ratio of the number of received packets $\times$ the size of a packet and the time taken to send all of the packets. Due to the difficulty of synchronizing the clocks on two machines, we measure the throughput for only the publisher. However, we also consider packet loss since the DDS protocol has a `UDP-like' property.  That is, it does not guarantee packet delivery since the data on this pub/sub system is designed for multicast real-time transmission.
\end{enumerate}

 
To avoid packet loss caused by the traffic congestion, we have included a `cooling-off' period after sending a packet. For example, in Algorithm \ref{TPub}, we include a $1 ms$ sleep cycle for each transmission to avoid congestion. Setting this value is critical to the measurement of throughput since a long cooling-off period will guarantee almost 100\% delivery of packets, but will cause the throughput to decrease. A short cooling-off period will also decrease the throughput as more packets will be dropped. We set the cooling-off period at 1 ms based on our experimental observation. With this cooling-off period, the small packets (16 bytes) are almost always delivered successfully, whilst the larger packets are dropped more frequently, but not often. Moreover, to measure the exact throughput, the publisher has to know the number of packets that are successfully delivered to the subscriber. A publisher sending ``START" and ``DONE" messages allows a subscriber to count the number of packets successfully subscribed as shown in Algorithms \ref{TPub} and \ref{TSub}.

\begin{algorithm}[t]
	\caption{Latency (Publish)}\label{LPub}
	\begin{algorithmic}[1]
		\Procedure{Latency\_pub}{$m\_size$, $topic_1$, $topic_2$}
        \State Generate an $m\_size$ sized random message $msg$ 
		\State $T_1 = current\_time()$
		\State Publish $msg$ on $topic_1$ 
		\State Subscribe $msg_r$ on $topic_2$
		\If {$msg$ == $msg_r$}
		\State	$T_2 = current\_time()$
		\EndIf
		\State \Return $latency = (T_2 - T_1)/2$
		\EndProcedure
	\end{algorithmic}
\end{algorithm}

\begin{algorithm}[t]
	\caption{Latency (Subscribe)}\label{LSub}
	\begin{algorithmic}[1]
		\Procedure{Latency\_sub}{$topic_1$, $topic_2$}
		\While {TRUE}
		\State Subscribe $msg$ on $topic_1$
		\State Publish $msg$ on $topic_2$ 
		\EndWhile		
		\EndProcedure
	\end{algorithmic}
\end{algorithm}

\subsection{Performance Benchmark Scenarios}

We test the latency and throughput performance on two network settings:  
\begin{itemize}
	\item \underline{Wired}: We connect two machines directly using an Ethernet cable. Each ROS 2 implementation is on a virtual machine in VirtualBox. We use a ``Bridged adapter", attached to the Ethernet Network Interface Card (NIC), to allow for the network connections. 
	\item \underline{Wireless}: We connect two machines wirelessly without using an extra access point. For the connection, we set \textbf{MPub} as an access point and \textbf{MSub} as a device connected to \textbf{MPub} using the ``Mobile Hostpot" function. We could not use the ad-hoc protocol since this feature is deleted in Windows 10. We turned on ``Network discovery" to allow the bidirectional communication in the Windows network setting. ``Bridged adapters" to the wireless NICs are used for both machines.    
\end{itemize}

We estimate the performances using various forms of communication security, which  cover most of the remote access situations:

\begin{enumerate} 
\item \textbf{No Security}: In this scenario, ROS 2 and its publish and subscribe system (i.e., DDS system) do not use any security functions to protect communications. Therefore, all communications between the nodes in two different machines are \textit{plain text} and can be monitored by other nodes or adversaries.
\item \textbf{Cryptographic Algorithms}: In this scenario, Secure ROS 2 (SROS2) \cite{SROS2016} is used to encrypt its communication channel as guided by OMG \cite{DDS1_1}. SROS is a package that provides the tools and instructions to use ROS 2 on top of DDS-Security (i.e., any DDS middleware). The intent of SROS2 is to provide a reasonable security infrastructure while minimizing user disruptions such as computational overhead and API breakage (note that SROS1 is proposed to be included in the ROS 1 API \cite{SROS1}). Accordingly, SROS2 with eProsima's Fast RTPS are evaluated to measure the overhead by encrypting/decrypting communication traffic.
\item \textbf{SSL/TLS}: In this scenario, an encrypted channel between two nodes is established in advance through OpenVPN. The setting similar to Scenario 1 (No Security), but the communication is secured using VPN technology. The VPN connection  uses ``AES-128-CBC” for ``cipher" and ``SHA256” for ``auth" options. 

\end{enumerate}

\begin{algorithm}[t]
	\caption{Throughput (Publish)} \label{TPub}
	\begin{algorithmic}[1]
		\Procedure{Throughput\_pub}{$s$, $d$, $topic_1$, $topic_2$}
		\State $throughput = 0$
		\State Generate an $s$ sized random message $msg$ 
		\State Publish ``START" on $topic_1$ and Sleep($100ms$)
		\State $clock = current\_time()$ and $i=0$
		\While {$i < d$} 
		\State Publish $msg$ on $topic_1$ 
		\State $i = i +1$ and Sleep($1ms$)
		\EndWhile
		\State $clock = clock - current\_time()$
		\State Sleep($100ms$) and Publish ``DONE" on $topic_1$
		\State Subscribe to $result$ on $topic_2$
		\State $throughput$ = $calculate\_bps(s,res,clock)$
		\State \Return $throughput$
		\EndProcedure
	\end{algorithmic}
\end{algorithm}

\begin{algorithm}[t]
	\caption{Throughput (Subscribe)}\label{TSub}
	\begin{algorithmic}[1]
		\Procedure{Throughput\_sub}{$topic_1$, $topic_2$}
		\State $rPKT = 0$
		\While {TRUE}
		\State Subscribe $msg$ on $topic_1$
		\If {$msg$ == ``Start"} 
		\State $rPKT = 0$
		\ElsIf {$msg$ == ``Done"} 
		\State Publish $rPKT$ on $topic_2$ 
		\State $rPKT = 0$ 
		\Else
		\State $rPKT = rPKT + 1$ 
		\EndIf 
		\EndWhile
		\EndProcedure
	\end{algorithmic}
\end{algorithm}

\section{Benchmark Result and Analysis}\label{physical_analysis}
In the first experiment, the two nodes are connected via an Ethernet cable, while in the second experiment they are connected using Wi-Fi. The ways in which the estimated latency and estimated throughput is measured are identical in these scenarios. For the latency test, we generated various sized packets, from 16 bytes to 12,000 bytes, and repeated each test 1,000 times to obtain the estimated latency. The throughput is measured in a similar way to that of latency over varying packet sizes. We sent 100 packets per test and repeated the tests 100 times for the estimation.

\subsection{Performance on Wired Network}

\subsubsection{Estimated Latency}
\begin{figure}[t]
		\centering
		\subfloat[Wired Network\label{fig1}]{\centering
		\includegraphics[width=0.225\textwidth]{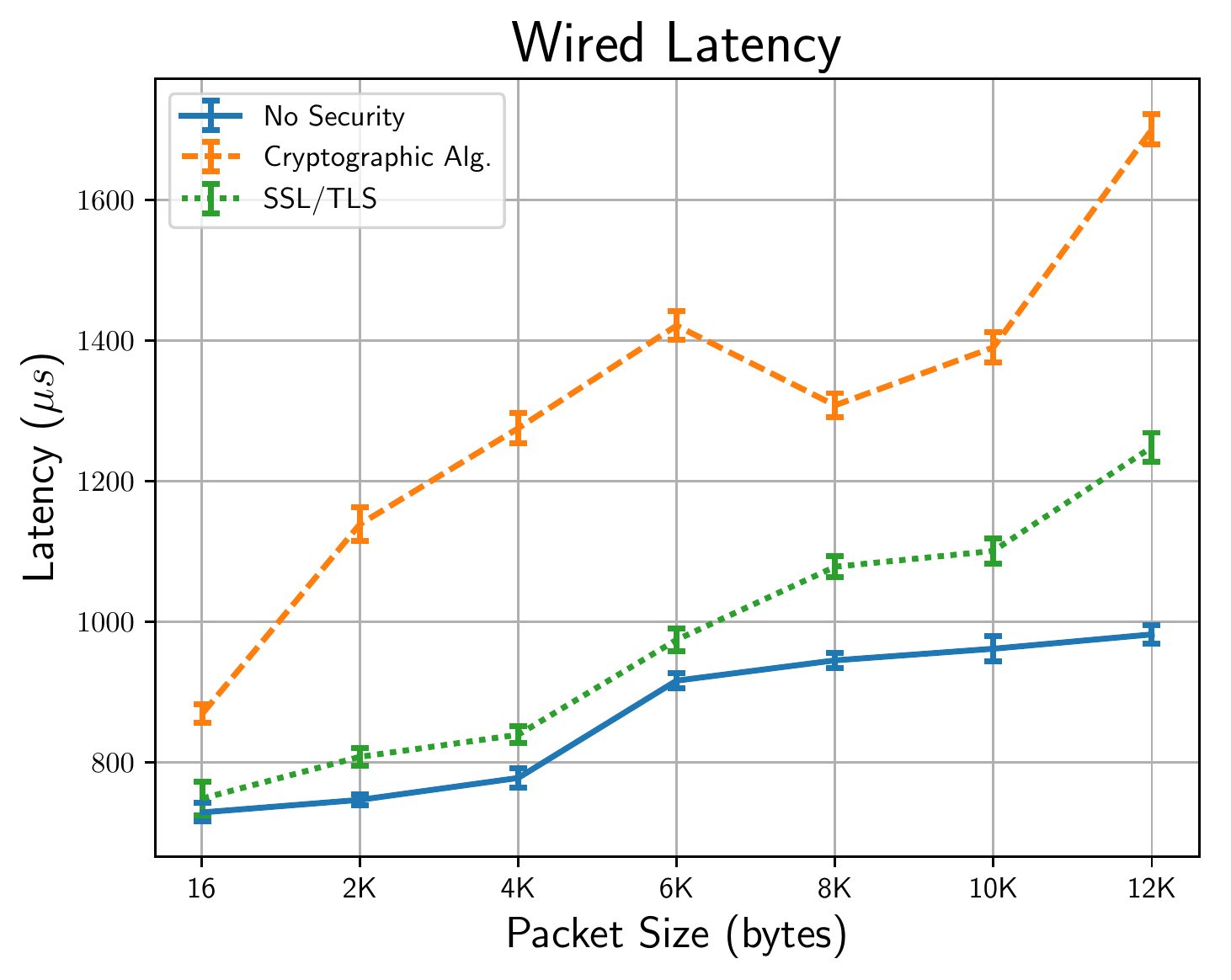} 
		}\hfill
		\subfloat[Wireless Network\label{fig6}]{\centering
		\includegraphics[width=0.225\textwidth]{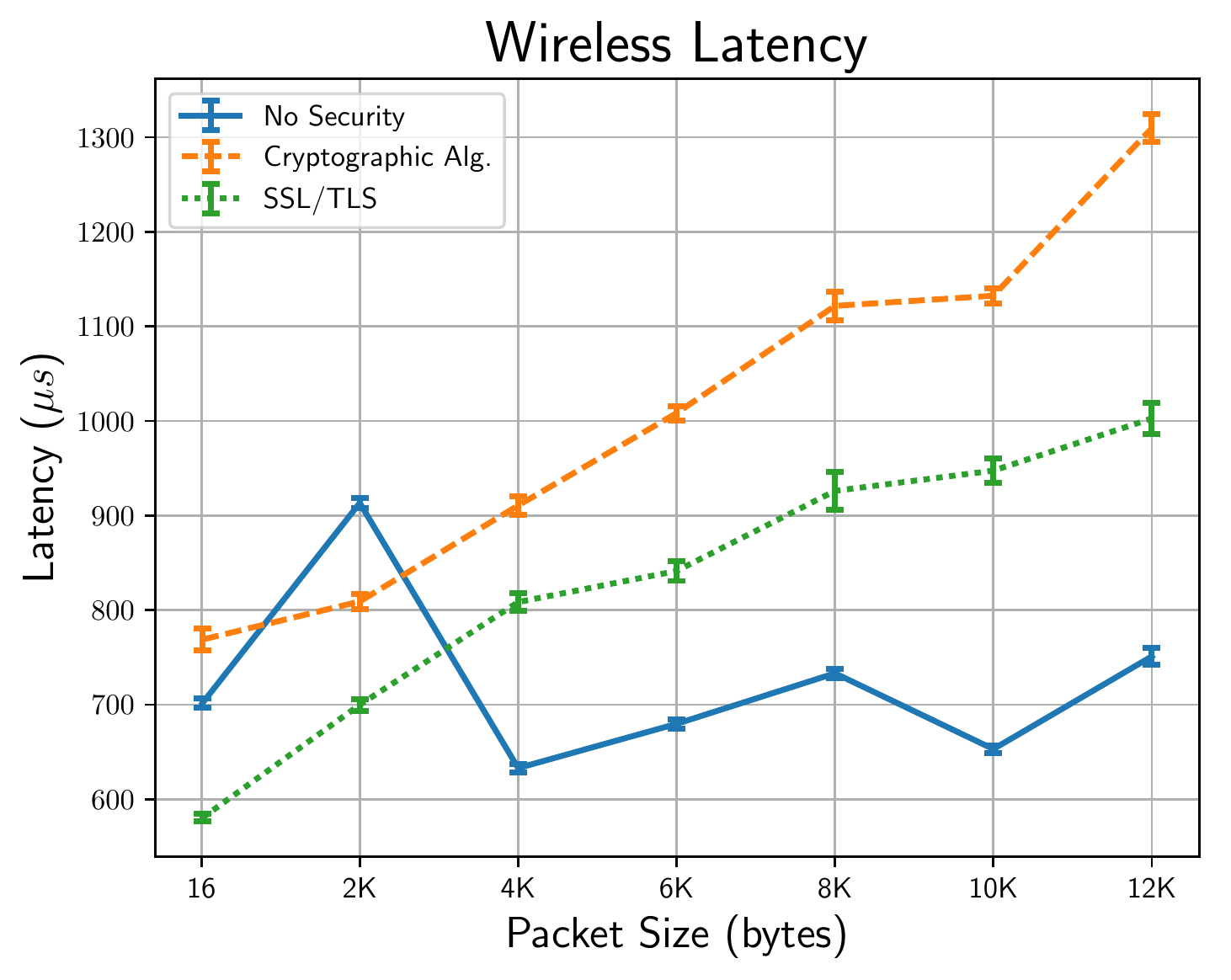} 
		}
		\caption{Latency per packet size in wired and wireless networks.}\vspace{-3mm}
\end{figure}

Figure \ref{fig1} displays the estimated latency for our scenarios. The DDS service (i.e., eProsima Fast RTPS) shows a low latency even when two nodes are remotely connected. Also, the latency is almost independent to the size of the packet if it is not encrypted. However, the encryption causes a delay that increases with the size of the packet. The most dramatic change is observed when using SROS2, where the latency is almost doubled in some instances. The use of a VPN can reduce this by more than half in the best case. 

\subsubsection{Estimated Throughput}
Figures \ref{fig4} and \ref{fig5} show the throughput of the ROS 2 application in a wired network. The figures show that the encryption in a ROS 2 application using SROS2 may cause a significant loss in throughput when compared to a connection with no security. It can also be seen that the VPN connection significantly outperforms SROS2 in most cases, and this performance gap can increase with packet size. 




\subsection{Performance on Wireless Network}

\subsubsection{Estimated Latency}
Figure \ref{fig6} shows the results from latency testing on a wireless network. We observed that using SROS2 can cause a significant delay in the network that increases with packet size, whilst the use of a VPN merely affects the latency of the packet delivery.


\begin{figure}[t]
		\subfloat[Wired Network\label{fig4}]{\centering
		\includegraphics[width=0.225\textwidth]{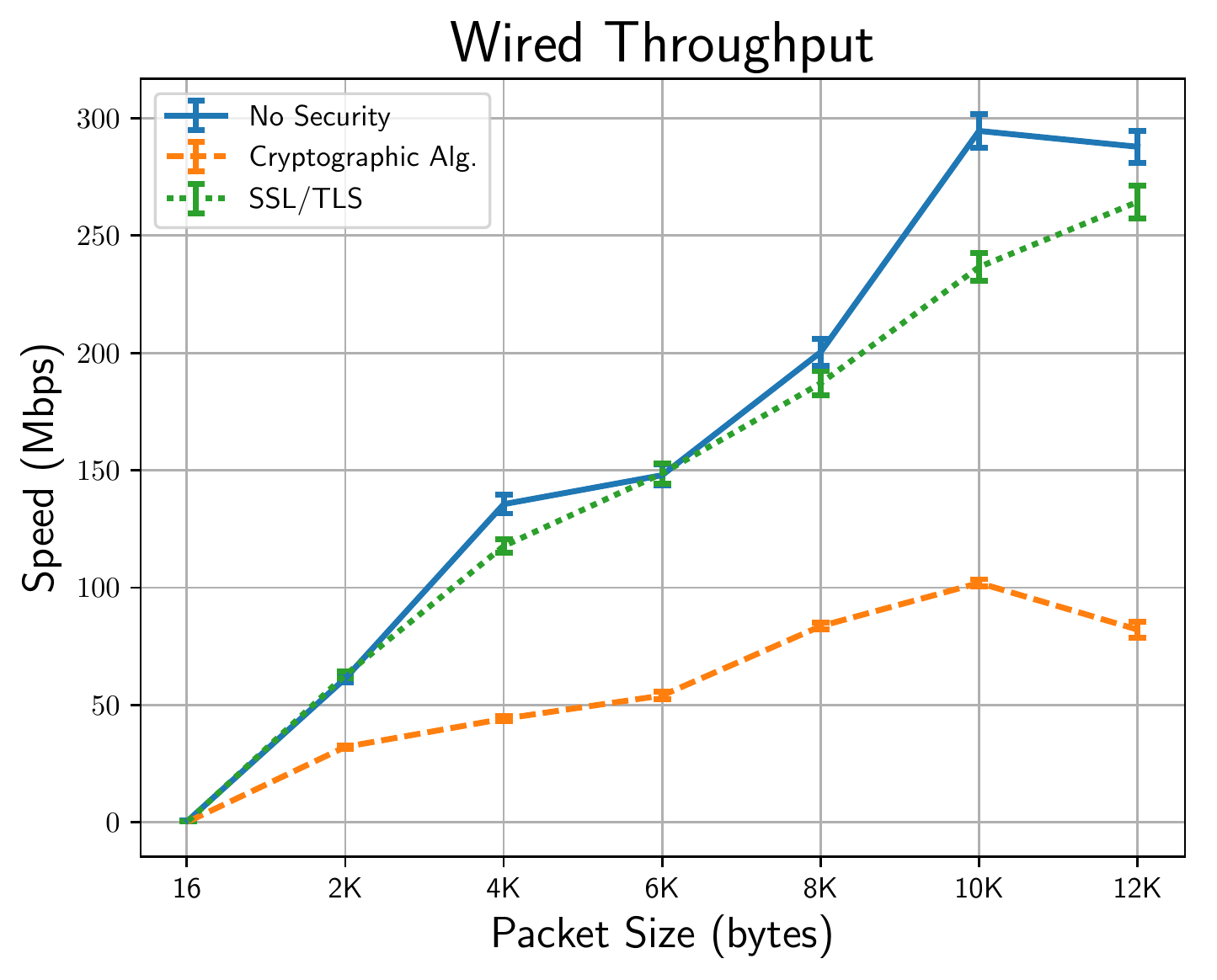} 
		} \hfill
		\subfloat[Wireless Network\label{fig8}]{\centering
		\includegraphics[width=0.225\textwidth]{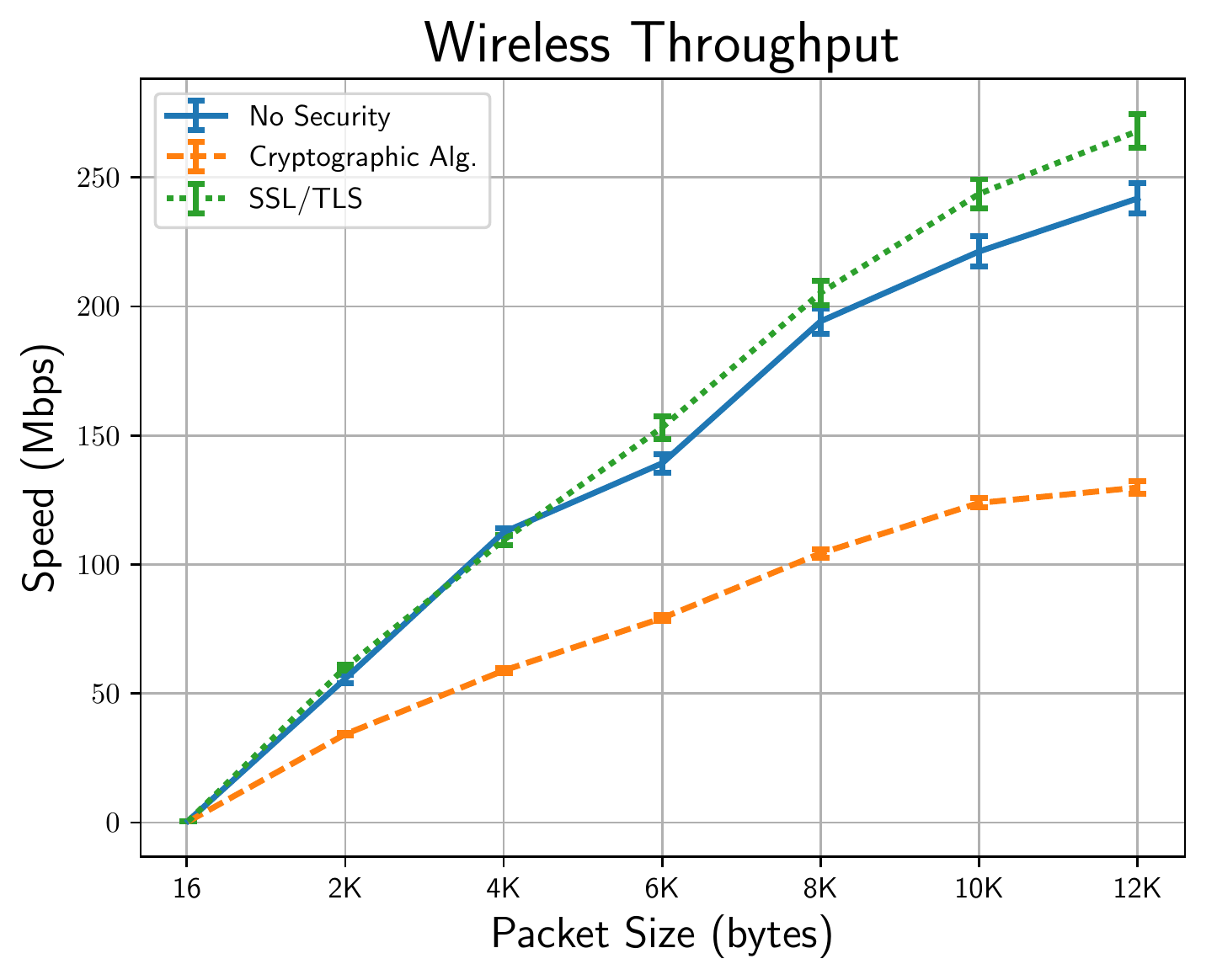} 
		}
		\caption{Throughput per packet size in wired and wireless networks.}
\end{figure}

\begin{figure}[t]
		\subfloat[Wired Network\label{fig5}]{\centering
		\includegraphics[width=0.225\textwidth]{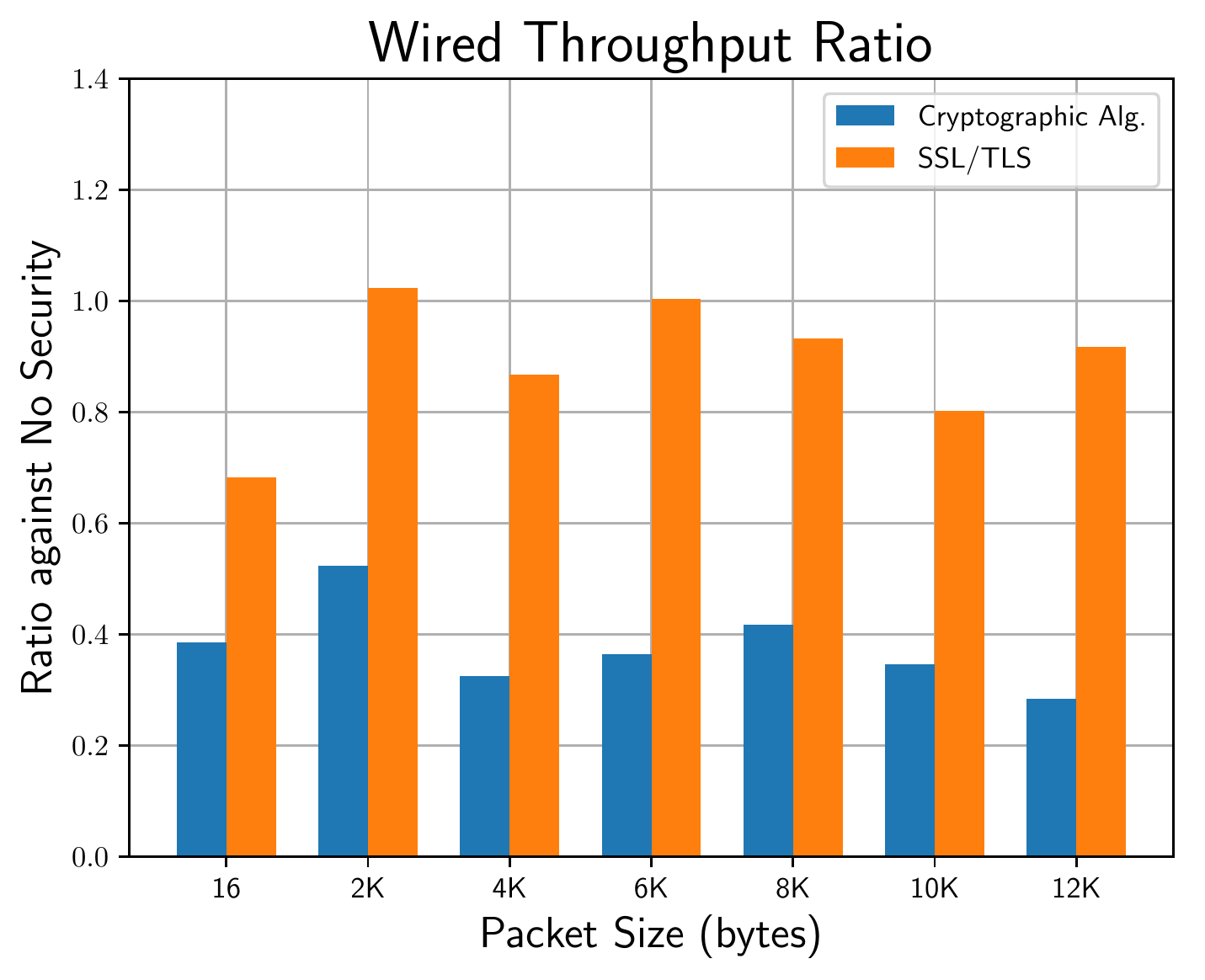} 
		} \hfill
		\subfloat[Wireless Network\label{fig9}]{\centering
		\includegraphics[width=0.225\textwidth]{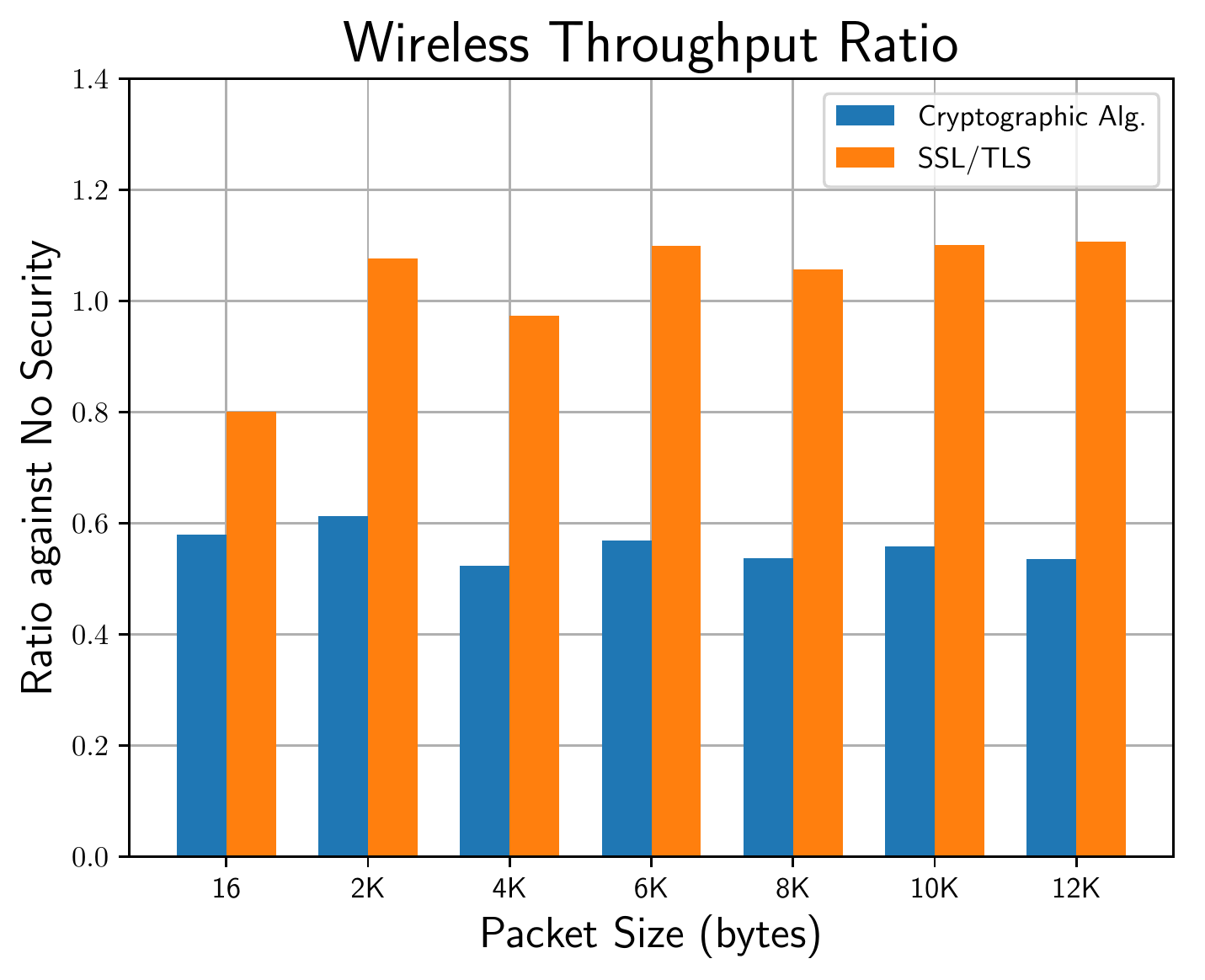} 
		}
		\caption{Throughput ratio against \textit{No Security} in wired and wireless networks. Here the y-axis shows the ratio between using cryptographic algorithms and SSL/TLS to secure the communication with the throughput from no security.}
\end{figure}

\subsubsection{Estimated Throughput}
Figure \ref{fig8} shows the throughput of the ROS 2 application in a wireless network. Similar to the latency results, we observed that using a VPN does not seriously affect the throughput in DDS, however using SROS2 causes a significant delay. 


Figure \ref{fig9} shows the loss of throughput caused by a VPN and SROS2 when compared to the no security scenario. The overhead associated with the use of a VPN vanished by end of the cooling-off period between packets sent (any loss is relatively small). Similar to the wired network, the overhead associated with the use of cryptographich algorithms reduces the throughput almost by half.




\subsection{Discussion}
Our results show that using a VPN is more efficient when two machines are connected, securely. It should be noted that the experiments only look at the connection between two machines, which may be used for emulating the communication between a controller and a CPS or between two CPSs. Unlike the DDS security protocol, a traditional VPN system has a server and client structure. All clients are connected to a server for the secure communication and therefore, the traffic is centralized to the VPN server. The DDS Security protocol does not have a server, but is instead designed for the secure connection between multiple distributed nodes. So it provides more decentralized features than a VPN. Each node has equal privilege, therefore the security communication cannot be disabled by compromising only a single machine in the system. Accordingly, SROS can be effective at providing communication security and resilience over multiple ROS 2 nodes when the use of a VPN is not desirable.



  
\section{Further Security Analysis in Implementation}\label{ddsrec}

In addition to the benchmark results on DDS and VPN security, we consider how the security functions of ROS 2 are properly implemented. Although the ROS community has released the non-beta version of ROS 2, the implementation of DDS Security of ROS 2 is still not mature. We performed a static code analysis on the DDS security implementation of ROS 2 (Ardent Apalone), particularly on eProsima's RTPS (the default installed with ROS 2). We found some notable problems in DDS Security specification and implementation. 

\subsection{Forward Secrecy}
In the DDS protocol, only two types of algorithms, ``DH+MODP-2048-256'' and ``ECDH+prime256v1-CEUM'', are supported. 
These handshake algorithms (DH and ECDH) do not support forward secrecy (FS). In cryptography, FS is a property of secure communication protocols in which the compromise of any long-term keys also compromises past session keys. Since the DH public key is fixed in the certificate and its corresponding private key is also fixed, all past session keys can be computed by an adversary if a long-term private key is compromised and all past traffic was recorded by the adversary. 

An easy way to support FS is to not use DH parameters as a long-term key. This is usually denoted as ``DHE'' in the SSL and TLS protocol, where DH parameters are generated for every new connection. The adversary then cannot compute the key used in past communications. Applying this protocol does not change a program much because DDS already supports all cryptographic algorithms that the change requires, but adopting this protocol certainly improves ROS 2 security. 
This lack of stronger security option does not belong to the specific eProsima implementation, rather, the deficiency exists in the OMG DDS security specification \cite{DDS1_1}.

\subsection{Static Code Analysis}

We used Cppcheck \cite{CPPCHECK} as a static analysis tool to report any notable errors within the ROS 2 Ardent Apalone source code. The resulting report detailed 19 errors with 12 false positives and only the remaining seven errors requiring closer observation to be fixed.
\begin{itemize}
\item Two errors are found in the eProsima RTPS. One error is caused by skipping initialization of a variable while the other is caused by skipping the freeing of a variable.
\item An error is found in the ROS Middleware Interface (RMW) which potentially prints out a NULL pointer. 
\item Four errors are found in RViz which relate to problems in memory management and can potentially cause a memory leakage.
\end{itemize}
While the specifics of the errors are not particularly notable, the process of employing a static code analysis tool is an important part of verifying the security of a CPS. Recently, DeMarinis et al. \cite{2018arXiv180803322D} showed that they were able to find over 100 publicly-accessible hosts running ROS while scanning for open IP addresses. They even unintentionally found two of their own robots as part of the scan (one Baxter robot and one drone). Similar to their study, this study echos the necessity of understanding the security settings when deploying a robot so that an adversary cannot access or control a robot capable of being remotely moved in ways dangerous both to the robot and the objects around it.

\subsection{Inconsistency between DDS Specification and RTPS}

We found two main inconsistencies between the DDS specification \cite{DDS1_1} and its implementation:
\begin{enumerate}
    \item There is a mandated 96 bit sized initial vector (IV) to be used in the implementation of AES-GCM (leaving the right-most 32 bits for a counter), while the actual size in RTPS is 128 bits. Some implementations may not accept this parameterization by default.
    \item A random number generator to generate challenges for master keys must be compliant with the NIST recommendation \cite{Barker:2012:SRR:2206302}. However, we observed that the current implementation relies only on the \texttt{BN\_rand} function that is provided by the OpenSSL library and does not comply with \cite{Barker:2012:SRR:2206302}.
\end{enumerate}

\subsection{Discussion}

We observed that the eProsima RTPS relies highly on the OpenSSL library. The OpenSSL community already provides a reliably secure version of the library, which is the OpenSSL FIPS objective module. Replacing the general OpenSSL library with the FIPS validated module may greatly improve the security of the current implementation and also provide the security functions which is currently deficient such as zeroization of secret data.

We also formally verified DDS handshake protocol using ProVerif \cite{DBLP:journals/jlp/BlanchetAF08}, which is automated protocol verifier. We could not find any problems on the protocol. It means that the correctness of the security protocol is verifiable.


\section{Conclusion}

In this paper, we discuss the ROS 2 security features and evaluate them in terms of impact on performance. We discussed two possible settings that ROS 2 can use to protect their communication in practice and measure their effects on latency and throughput performances. Particularly, we build our own ROS 2 applications to test those parameters over various settings by combining wired and wireless networks. 

Our results show that using a reliable VPN application is a better choice so far if an application of ROS 2 requires a simple system architecture, such as a CPS with a controller or a centralized server-client architecture. In addition to the poor performance when compared to the VPN, we also observe that the accuracy of the DDS security implementation to the specification should not be assumed as correct. After analyzing the default DDS middleware with ROS 2 we found that it did not conform to the security specification by OMG. This can cause a CPS to be left in vulnerable situations. 


ROS 2 is still under rapid development with updated versions being published every few months. Continuously monitoring the implementation of ROS 2 to check violations to the relevant security specifications and logical errors, which can be linked to potential vulnerabilities is greatly beneficial to enhance the security of ROS 2. Since ROS 2 is a software library, its applications may vary; therefore the heterogeneity of ROS 2 applications makes confirming the security of applications difficult. Developing the tools and procedures that enable developers to deploy during and after the development process will reduce the risk of security vulnerabilities of resulting CPS systems.

\section*{LEGAL}
DISTRIBUTION A. Approved for public release: distribution unlimited. 
This material is based upon work funded and supported by the Department of Defense under Contract No. FA5209-17-P-0007 with the Commonwealth Scientific and Industrial Research Organization (CSIRO).

\bibliographystyle{IEEEtran}
\bibliography{IEEEabrv,bibfile}

\addtolength{\textheight}{-12cm}   

\end{document}